\documentclass{aastex6}

\usepackage{graphicx}
\usepackage{amssymb}
\usepackage{epstopdf}
\usepackage{float}

\def\lae{\mathrel{<\kern-1.0em\lower0.9ex\hbox{$\sim$}}}  
\def\gae{\mathrel{>\kern-1.0em\lower0.9ex\hbox{$\sim$}}}  
\def\msun{\ifmmode{\ {\rm M}_\odot}\else{$ {\rm M}_\odot$}\fi}  
\def\msunyr{\ifmmode{\msun \ {\rm yr}^{-1}}\else{$\msun \ {\rm 
yr}^{-1}$}\fi}  

\fullcollaborationName{The Friends of AASTeX Collaboration}

\begin{document}


\title{A Mechanism for Stimulating AGN Feedback by Lifting Gas in Massive Galaxies}

%% Use \author, \affil, plus the \and command to format author and affiliation 
%% information.  If done correctly the peer review system will be able to
%% automatically put the author and affiliation information from the manuscript
%% and save the corresponding author the trouble of entering it by hand.
%%
%% The \affil should be used to document primary affiliations and the
%% \altaffil should be used for secondary affiliations, titles, or email.

%% Authors with the same affiliation can be grouped in a single
\author{B.R. McNamara\altaffilmark{1,2}, H.R. Russell\altaffilmark{3}, P.E.J. Nulsen\altaffilmark{4,5}, M.T. Hogan\altaffilmark{1,2}, A.C. Fabian\altaffilmark{3}, F. Pulido\altaffilmark{1}, A.C. Edge\altaffilmark{6}}
%\affil{University of Waterloo \\
%200 University Ave W, Waterloo \\
%ON N2L 3G1, CANADA}


\altaffiltext{1}{Department of Physics and Astronomy, University of Waterloo, 200 University Avenue West, Waterloo, ON N2L 3G1, Canada}
\altaffiltext{2}{Perimeter Institute for Theoretical Physics, Waterloo, ON N2L 2Y5, Canada}
\altaffiltext{3}{Institute of Astronomy, Madingley Road, Cambridge CB3 0HA, UK}
\altaffiltext{4}{Harvard-Smithsonian Center for Astrophysics, 60 Garden Street, Cambridge, MA 02138, USA}
\altaffiltext{5}{ICRAR, University of Western Australia, 35 Stirling Hwy, Crawley, WA 6009, Australia}
\altaffiltext{6}{Centre for Extragalactic Astronomy, Department of Physics, Durham University, Durham DH1 3LE, UK}

\begin{abstract}

Observation shows that nebular emission, molecular gas, and young stars in giant galaxies are associated with rising X-ray bubbles inflated by radio jets launched from nuclear black holes. We propose a model where molecular clouds condense from low entropy gas caught in the updraft of rising X-ray bubbles.  The low entropy gas becomes thermally unstable when it is lifted to an altitude where its cooling time is shorter than the time required to fall to its equilibrium location in the galaxy i.e., $t_c/t_{\rm I} \lesssim 1$.   The infall speed of a cloud is bounded by
the lesser of its free-fall and terminal speeds, so that the infall
time here can exceed the the free-fall time by a significant factor. This mechanism is motivated by ALMA observations revealing molecular clouds lying in the wakes of rising X-ray bubbles with velocities well below their free-fall speeds. Our mechanism would provide cold gas needed to fuel a feedback loop while stabilizing the atmosphere on larger scales. The observed cooling time threshold of $\sim 5\times 10^8~\rm yr$ --- the clear-cut signature of thermal instability and the onset of nebular emission and star formation--- may result from the limited ability of
radio bubbles to lift low entropy gas to altitudes where thermal instabilities can ensue.  Outflowing molecular clouds are unlikely to escape, but instead return to the central galaxy in a circulating flow. We contrast our mechanism to precipitation models where the minimum value of $t_c/t_{\rm ff} \lesssim 10$ triggers thermal
instability, which we find to be inconsistent with observation.
\end{abstract}

\keywords{galaxies: clusters: general--- galaxies: evolution--- galaxies: individual (M87, MS0735+7421,Abell 2029)---X-rays: galaxies: clusters }



\section{Introduction} \label{sec:intro}

Energetic feedback from nuclear black holes is thought to regulate the growth of massive galaxies from their nascency during the quasar era \citep{fabian12} through to their maturity 
as radio galaxies  \cite[][]{McNamara07,McNamara12}.
Atomic and molecular outflows observed in quasars and active galaxies apparently regulate star formation, and in some instances, sweep the host galaxy of its gas during the most active phases of galaxy growth \cite[e.g.,][]{nesvadba08, Feruglio10,Morganti05,Arav08,Alexander10}.  Mrk 231, for example, has revealed several $\times 10^8 ~\rm M_\odot$ of molecular gas flowing out of
its inner kpc or so with velocities exceeding $700 ~\rm km~s^{-1}$ \cite[][]{Rupke07}. The evolution of giant ellipticals at late times is governed instead by radio jets that heat the hot, X-ray atmospheres of galaxies and clusters that would otherwise cool and sustain star formation \cite[][]{Birzan04,Best07,Rafferty08,Dunn08}. Archetypes include NGC 1275 in Perseus \cite[][]{Fabian07}, M87 \cite[][]{Forman07,Forman16}, and the normal giant elliptical galaxies M84 \cite[][]{Finoguenov01}, and NGC 5813 \cite[][]{Randall15}.  This so-called radio mode or radio-mechanical feedback is responsible in whole or in part for the inefficiency of star formation in the central galaxies of massive halos as they age, leaving them red and dead \cite[][]{Croton06,Bower06}.  

New observations of molecular gas in central galaxies made with the Atacama Large Millimeter Array (ALMA) and new numerical simulations of radio bubbles rising in hot atmospheres suggest a richer, more complex picture. 
%ALMA's extraordinarily high resolution and sensitivity have provided surprising and important clues to the origin of thermal instability, cooling, and star formation in galaxies.
Following on discoveries of upwards of $10^9~\rm M_\odot$ of molecular gas in central cluster galaxies \cite[][]{Edge01,Salome03}, ALMA and IRAM observations of a half dozen or so central galaxies, including NGC 1275 \cite[][]{Salome11,Salome08a,Salome08b,Salome06},  Abell 1835 \cite[][]{McNamara14}, NGC5044 \cite[][]{David14}, and PKS 0745-191 \cite[][]{Russell16} indicate that molecular clouds are either lifted out by, or condensing along the trajectories of, buoyantly-rising X-ray bubbles inflated by radio jets. Furthermore, the molecular clouds are moving at surprisingly slow speeds with respect to the velocity dispersion of the stars \cite[][]{Russell16} and well below the escape speed of the central galaxy.  Observation indicates the molecular clouds are circulating in the potential well of the galaxy while fueling star formation at rates of several to several tens of solar masses per year \cite[][]{Salome11,McNamara14}.

Star formation and associated nebular emission are hallmarks of galaxies and clusters hosting hot atmospheres with cool cores \cite[][]{Johnstone87,Heckman89}.  Chandra X-ray observations have established a cooling time threshold in these systems for the onset of nebular emission and star formation. Star formation and nebular emission are prevalent when the central atmospheric cooling time falls below $\sim 5\times 10^8~\rm yr$, or similarly, the central entropy parameter falls below $30 \rm~ keV~cm^2$ \cite[][]{Rafferty08,Cavagnolo08}.  Systems lying above the central cooling time threshold are usually devoid of cooling gas and star formation, while those below usually are not.  The threshold may be related to the onset of thermal instability in the hot atmosphere \cite[][]{Voit08} but the reasons for its numerical value are not understood.

Theoretical studies of thermal instabilities in cluster atmospheres have attributed the cooling time threshold to thermal conduction, which tends to stabilize
cooling atmospheres, and to the ratio of the local cooling time to free fall time for thermally unstable clouds \cite[][]{McCourt12,Sharma12,Gaspari12,Voit15a}. These studies concluded that when the ratio of the cooling time to free-fall time falls below $t_c/t_{ff} \lesssim 10$, thermal instability ensues fueling nebular emission and star formation. While some systems are consistent with this criterion \cite[][]{McCourt12,Voit15a,Voit15b,Loubser16}, we show here that the observed values of this ratio are governed almost entirely by the cooling time (the numerator), not the free-fall time.  Furthermore, the criterion as applied in these studies 
forecasts H$\alpha$ emission less reliably than the central cooling time or central entropy alone. 
 
Motivated primarily by new ALMA observations, we suggest instead that thermal instabilities occur preferentially when cool, X-ray emitting gas lying within the central galaxy is lifted to higher altitudes behind buoyant X-ray bubbles inflated by radio AGN.  This effectively increases the {\em infall} time of the gas, promoting condensation into molecular clouds in the bubbles' wakes.  The surprisingly slow molecular cloud velocities found by ALMA \cite[][]{McNamara14,Russell16} indicate that the infall timescale ($t_I$) is substantially longer than the free-fall timescale, 
promoting thermal instability. In this new picture of feedback, rising X-ray bubbles responsible for heating  hot atmospheres and regulating cooling and star formation simultaneously promote cooling in their wakes, fuelling an ongoing feedback loop in a mechanism we refer to as stimulated feedback.

%However, the absence of strong emission lines from OVII and Fe XVII in clusters, expected in a radiative cooling flow (Peterson \& Fabian 2006), and  associated AGN activity indicate that cooling is stabilized by AGN feedback (McNamara \& Nulsen 2007, 2012, Fabian 2012).  Some central cluster and group galaxies are actively forming stars (McDonald et al. 2015),but only in clusters or groups where the central atmospheric (i.e.<12 kpc) radiative cooling time is shorter than $\sim 5\times 10^8~\rm yr$ or equivalently central entropy less than $30 \rm~ keV~cm^2$ (Rafferty et al. 2008).  This sharp threshold for star formation is likely related to the onset of thermal instability in the hot atmosphere, however its origin in poorly understood.  

\section{The Onset of Nebular Emission and Star Formation in central galaxies}

Despite a common misperception that central cluster galaxies are dormant, nebular emission, star formation, and other indications of cooling gas are common in cool core or cooling flow clusters \cite[][]{Cowie83,Hu85,Heckman89}.  The close association between cooling atmospheres and star formation directly links the growth of central galaxies and their nuclear black holes to the reservoir of hot gas surrounding them.  Chandra X-ray images have revealed thermodynamic thresholds indicating that nebular emission and star formation ensue when a
hot atmosphere's central cooling time or entropy index fall below $t_c \lae 5 \times 10^8~ \rm yr$ and $K \lae 30 \rm ~kev~cm^2$, respectively \cite[][]{Rafferty08,Cavagnolo08}.  Galaxies hosting atmospheres lying above these thresholds do not shine with nebular emission or star formation while those lying below usually do. The thresholds are remarkably sharp, indicating a direct connection between galaxy evolution, feedback, and atmospheric cooling.  
The thresholds forecast H$\alpha$ emission more reliably than star formation, most likely because small levels of molecular and atomic gas ($\lesssim 10^6 ~M_\odot$) emit detectable levels of H$\alpha$ emission before the galaxy has accumulated enough molecular gas ($\sim 10^9 ~M_\odot$) to fuel appreciable levels of star formation. The cooling time and entropy thresholds point to thermal instability in hot atmospheres fueling nebular emission and star formation \citep{nulsen86,Pizzolato05}. However, a convincing theoretical explanation of its value has proved elusive.

%These points are illustrated in two of the three examples discussed below (M87, MS0735+7421) shine in nebular emission but lack visible signs of star formation.  As expected, the central cooling times of both lie below the threshold. The third, Abell 2029, also falls below the cooling time threshold yet shows no sign of nebular emission or star formation for reasons discussed below. We can explain the coincidence between the H$\alpha$ and star formation thresholds because such systems are unstable to cooling but the surface density of molecular gas lies below required for gravitational collapse to stars. We do not understand why the threshold exists, although theoretical studies have attributed it to the onset of cooling thermal instabilities which are partially stabilized by conductive heating (Soker Voit et al. 2008).

\subsection{Observational Inconsistency with $t_c/t_{ff}$ Threshold}

Several studies have argued that the cooling time and entropy thresholds are a consequence of thermal instabilities that develop in hot atmospheres when the ratio of $t_c/t_{ff}$ falls below 10 \cite[][]{McCourt12,Sharma12,Voit15a}. If true, the ratio should more reliably forecast H$\alpha$ emission than the cooling time or entropy index alone.
We find that the $t_c/t_{ff}\lae 10$ criterion as applied in these studies is less reliable. For example, of more than 200 cluster cores studied by \citet[][]{Cavagnolo08} only five lying below the cooling time/entropy threshold failed to shine with H$\alpha$ emission, an iconic example being Abell 2029,  which we discuss in detail below.  In contrast,
only 10 of 43 systems in \citet{Voit15b} with detectable H$\alpha$ emission met the $t_c/t_{ff}\lae 10$ criterion.

% Will uncomment these once we have the figures...
\begin{figure}[htbp] %  figure placement: here, top, bottom, or page
	\figurenum{1}
	\centering
	\gridline{\fig{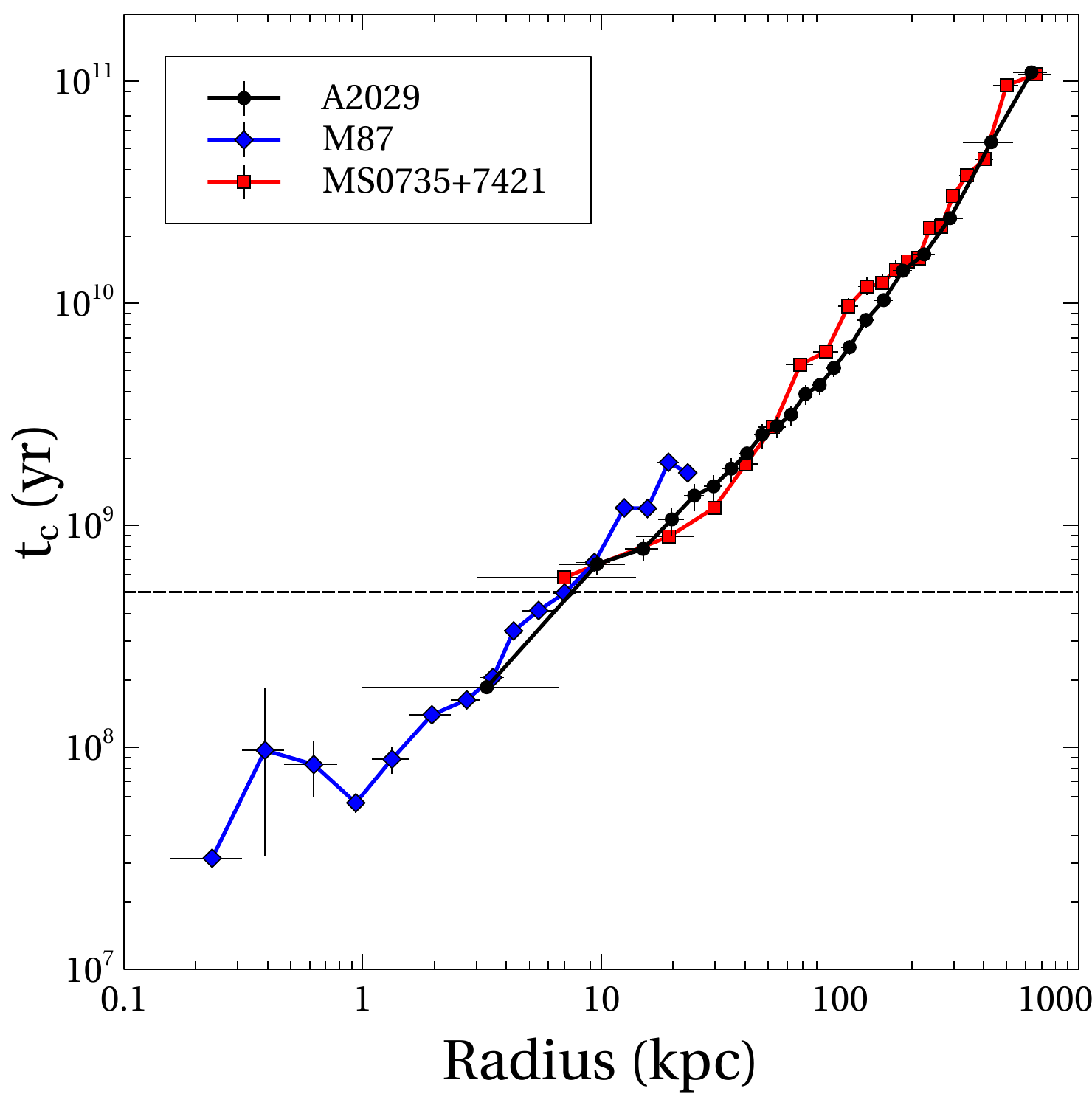}{0.3\textwidth}{}
          	  \fig{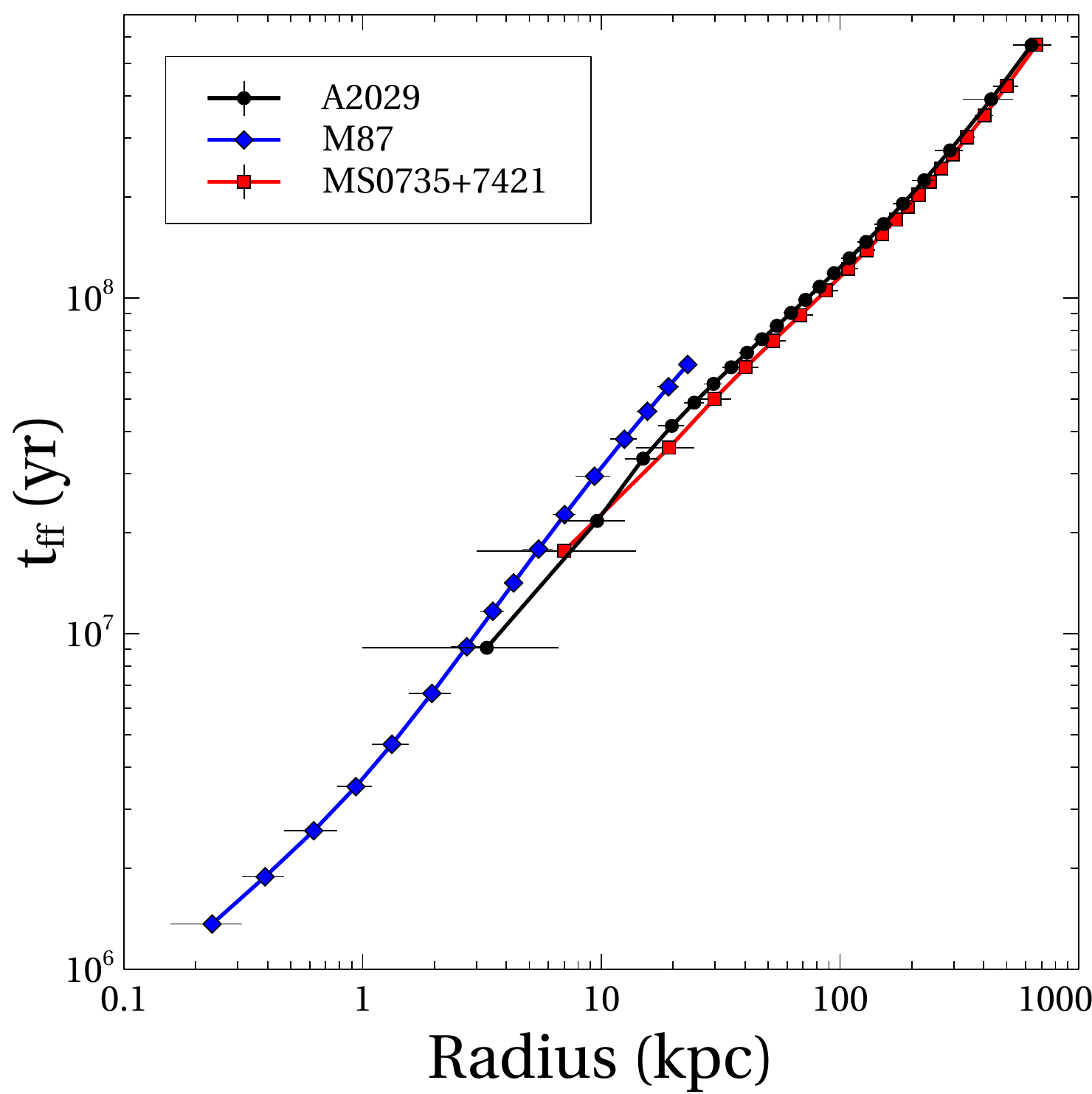}{0.3\textwidth}{}
              \fig{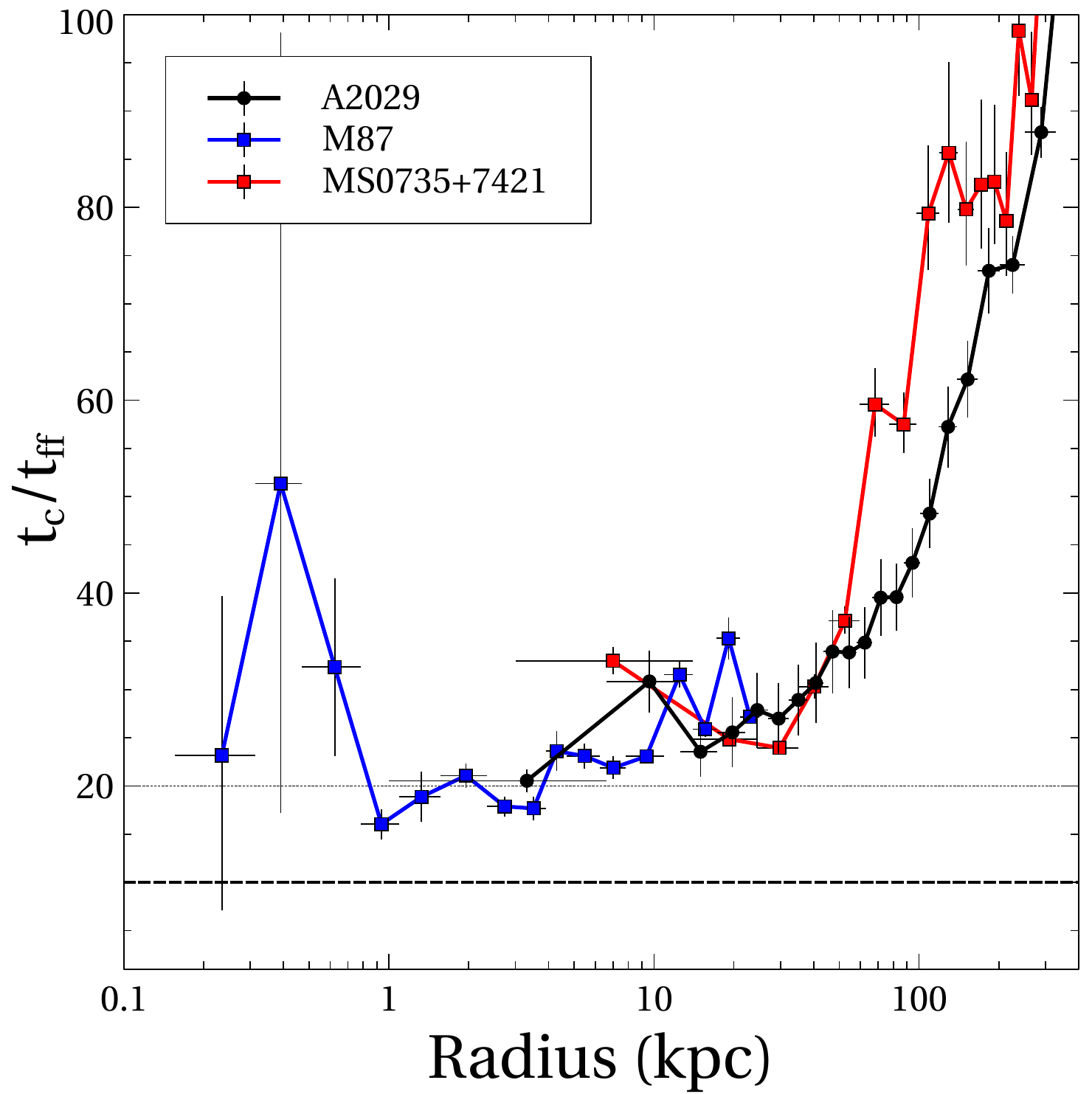}{0.3\textwidth}{}
    }
	\caption{Radial run of cooling time calculated using deprojected gas densities, free-fall times, and the ratio of these quantities.}
	\label{fig:f1}
\end{figure}

%\subsection{Is Thermal Instability Driven by the Cooling Time or the Ratio of Cooling Time to Free-Fall Time?}

We illustrate this point by calculating the $t_c/t_{ff}$ threshold as it applies to three iconic central cluster galaxies for which we plot the cooling time, free-fall time, and their ratio in Figure \ref{fig:f1}. 
The center and right panels of Figure \ref{fig:f1} show the free-fall time profiles and $t_c/t_{ff}$ profiles, respectively, for each cluster. 
The cooling time profiles were derived from deprojected gas density profiles, which removes emission from hot gas at large atmospheric distances seen in projection.
The free-fall times were estimated for Abell 2029 and MS0735+7421 assuming hydrostatic NFW profile fits to the X-ray data beyond an altitude of about 30 kpc.  Those profiles were then grafted to isothermal profiles within 30 kpc whose masses are anchored to the stellar mass, which matches their stellar velocity dispersion profiles.  The M87 data were taken from \citet[][]{Russell15} who adopted the NFW2 profile from \citet[][]{Romanowsky01}.  A complete discussion of our methodology and results for a large sample of clusters will be presented in Hogan {\em et al.} and Pulido et al. {\em et al.} in preparation. 

We use these three iconic examples to illustrate a general trend. The central cooling times for 
all three objects fall near to or below the cooling time threshold of $\sim 5\times 10^8$ yr, indicated by a dashed line in the
left panel of  Figure \ref{fig:f1}.  M87 \cite[][]{Sparks04} and MS0735+7421 \cite[][]{Donahue92} shine in nebular  emission within $5-30$ kpc of their nuclei, as expected based on their short central cooling times.  H$\alpha$ emission in central galaxies is associated with the presence of molecular clouds  \cite[][]{Edge01}.  The third cluster, Abell 2029, has revealed no appreciable H$\alpha$ emission \cite[][]{mcdonald10}, [O II]$\lambda$3727\AA ~emission, or star formation \cite[][]{McNamara89} in its central galaxy, despite falling well below the cooling time threshold \cite[][]{Cavagnolo08,Rafferty08}.  All three lie well
above $t_c/t_{ff}\lae 10$. By this criterion, all three should not shine in H$\alpha$ emission, yet two do.  At the same time, the cooling time threshold predicts
all three should shine with H$\alpha$ emission, yet Abell 2029 does not.   Something is awry, and we suggest new and interesting
physics is needed to solve the problem.  A more detailed description of these objects is given in the Appendix.

Regardless of hosting AGN spanning four decades of radio-mechanical power, the cooling profiles for all three are remarkably similar.  Due to their larger distances, the profiles for Abell 2029 ($z=0.077$) and MS0735+7421 ($z=0.216$), are unresolved below a few kpc. Nevertheless, M87's cooling times at 2 and 6 kpc are similar to Abell 2029 and MS0735+7421 at similar altitudes. 
Based on their similar shapes it would be tempting to suggest the cooling time profiles for Abell 2029 and MS0735+7421 continue to decline into the nucleus despite the vast differences in AGN mechanical power between the three objects.  

The key point is, despite short central cooling times, the ratios of cooling time to free-fall time all lie well above 10.  Therefore, all should be thermally stable and devoid of H$\alpha$ emission, yet two are not. \citet[][]{McCourt12} suggested that an upward departure from the $t_c/t_{ff}\lesssim 10$ criterion in systems with bright H$\alpha$ emission, such as 
M87 and MS0735+7421, may be a consequence of a temporary decrease in central gas density in response to AGN heating.   This explanation cannot be excluded out of hand given the scatter in the
central cooling time profiles of clusters \citep{Panagoulia14}.
%Observations show that cooling time profiles vary in amplitude as a function of radial position (altitude) by a factor of about 10 between altitudes of $10-100$ kpc (Panagoulia et al. 2014), so hysteresis could be a factor affecting the $t_c/t_{ff}$ profiles.  However, it is not clear that AGN are driving the variation rather than the converse (Cavagnolo et al. 2008, Rafferty et al. 2008), and the dispersion in cooling profiles beyond 100 kpc suggests that other factors, such as gravity rather than AGN cause the dispersion.
However, that the cooling profiles in Figure 1 are so similar, despite ongoing AGN activity in both M87 and MS0735+7421, indicates that a dramatic increase in AGN power need not lead to a dramatic upward response in either atmospheric cooling time, the radial run of $t_c/t_{ff}$, or central gas density.  This point is further illustrated in Figure 2, where we plot the radial run of gas density in the hot atmospheres of the three objects discussed here.  In addition, we have included  Abell 1835, which hosts one of the largest reservoirs of molecular gas ($5\times 10^{10}~\rm M_\odot$) and one of the highest
star formation rates ($\sim 200  ~\rm M_\odot ~yr^{-1}$) known \citep{McNamara14, mcnamara06}.  Figures 1 and 2 illustrate the surprisingly small variation in both gas
density and cooling time of the hot atmosphere, despite an enormous range of AGN power.   The objects shown in Figure 2 span five decades in AGN energy and nearly four decades in molecular gas mass.  Yet the variation in their gas densities at 10 kpc, which is where most studies find a minimum value in $t_c/t_{ff}$, lie in the narrow range of $2-10\times 10^{-2}~\rm cm^{-3}$.  Furthermore,
their cooling times at 10 kpc are nearly identical.   Despite vast differences in their star formation rates,  molecular gas masses, and most importantly, their AGN power, their central X-ray gas densities and cooling times are remarkably steady showing little evidence of cycling indicated in precipitation models.  Thus AGN contribute little to the scatter in central cooling times found by \citet{Panagoulia14}, but
instead probably reflect variations in their halo masses.   
%This inelastic response to AGN activity is consistent with the tendency for atmospheric entropy and cooling profiles to decline into the nuclei of central galaxies once resolution effects are accounted for \cite[][]{Panagoulia14}. 
The resilience of the central gas density and central cooling time to powerful AGN outbursts is, in fact, a key feature of hot atmospheres stabilized by continual and gentle AGN feedback \cite[][]{McNamara12}.  
%Therefore, departures above $t_c/t_{ff}=10$ in thermally unstable systems are unlikely to result from hysteresis associated with the cooling cycle.  Note that only the cooling time would be affected by hysteresis so that the cooling time/entropy threshold would be similarly affected yet it is not. 

\begin{figure}[htbp] %  figure placement: here, top, bottom, or page
   \figurenum{2}
   \centering
	\gridline{\fig{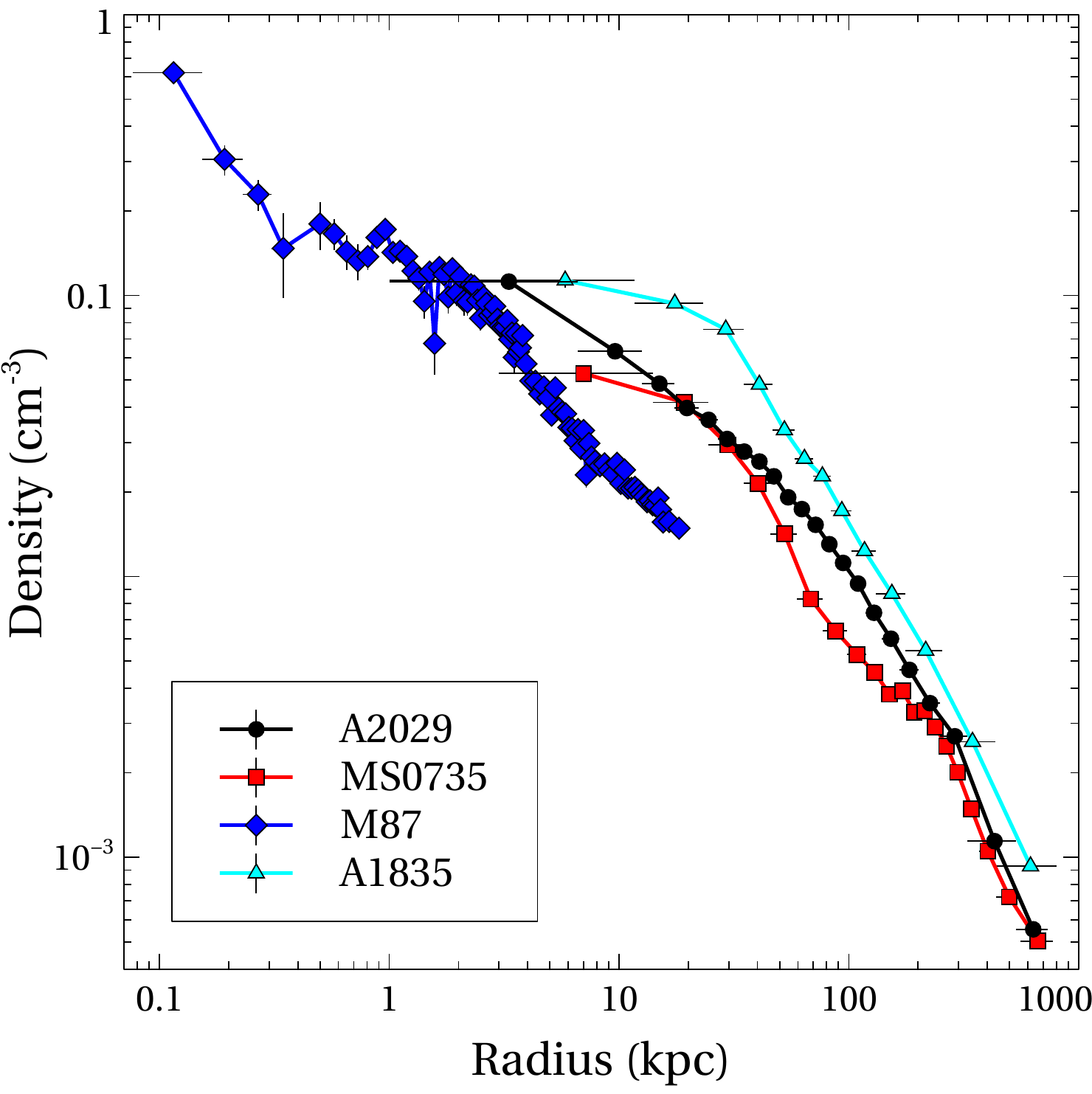}{0.5\textwidth}{}
            }
 \caption{Radial dependence of electron density for the three objects presented in Figure 1.  We have included Abell 1835 as an example galaxy with
 a large molecular gas mass $5\times 10^{10}\msun$ and a star formation rate approaching $200\msunyr$ \citep{mcnamara06}.  The outburst
 energies determined from cavity and shock front measurements are: M87, $5\times 10^{57} ~\rm erg$ (Forman et al. 2016), MS0735+7421, $9\times 10^{61} ~\rm erg$ \citep{Vantyghem14},
 Abell 1835, $4\times 10^{59} ~\rm erg$ \citep{mcnamara06}, and Abell 2029, no detectible shocks or cavities.  Despite having experienced AGN outbursts spanning five decades
 in energy over the past several tens of Myr, their gas densities at 10 kpc vary by only a factor of 5 and their cooling times at 10 kpc are nearly identical.}
   \label{fig:f2}
   \end{figure}

\begin{figure}[htbp] %  figure placement: here, top, bottom, or page
	\figurenum{3}
	\centering
	\gridline{\fig{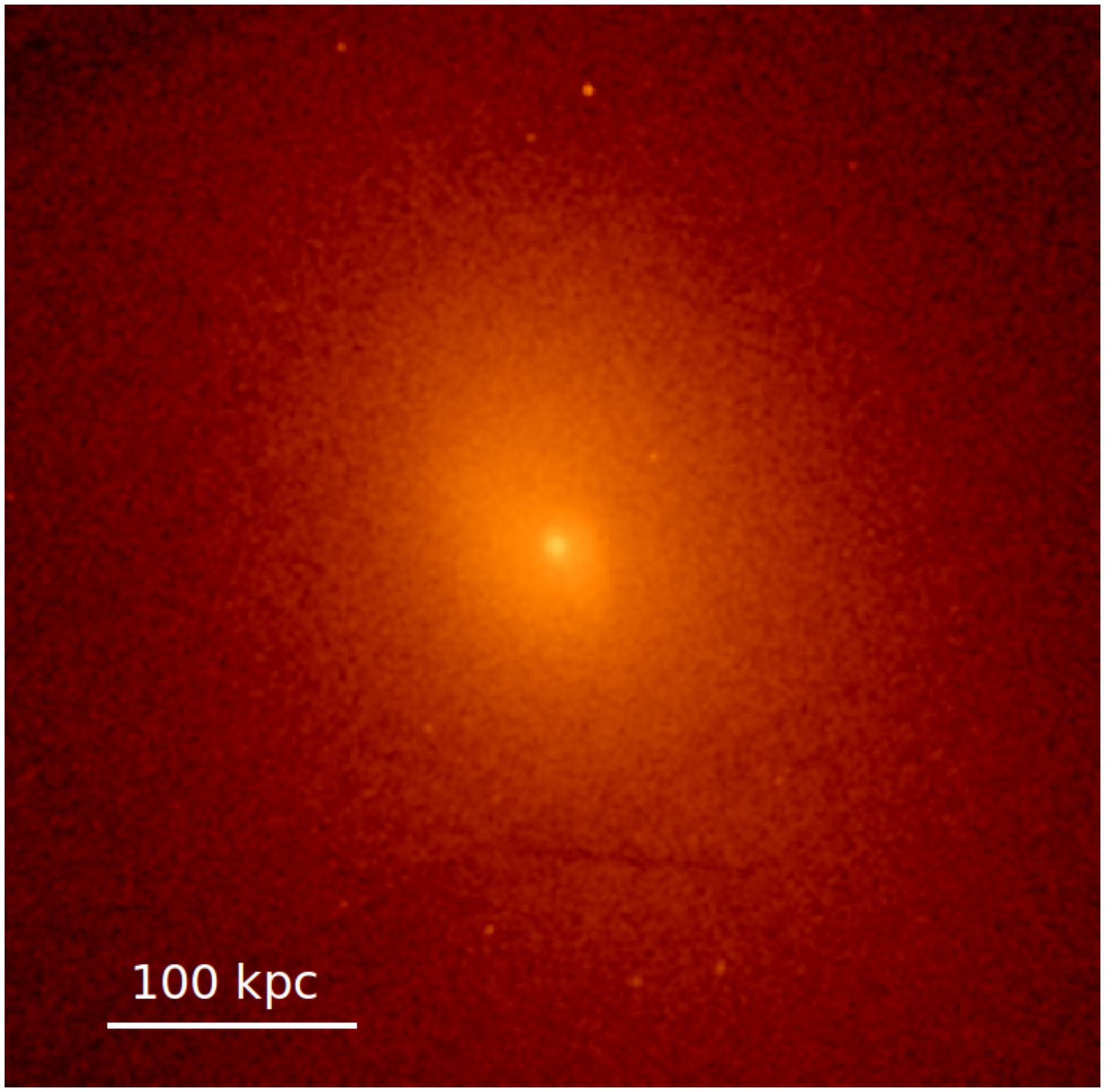}{0.3\textwidth}{Abell 2029}
          	  \fig{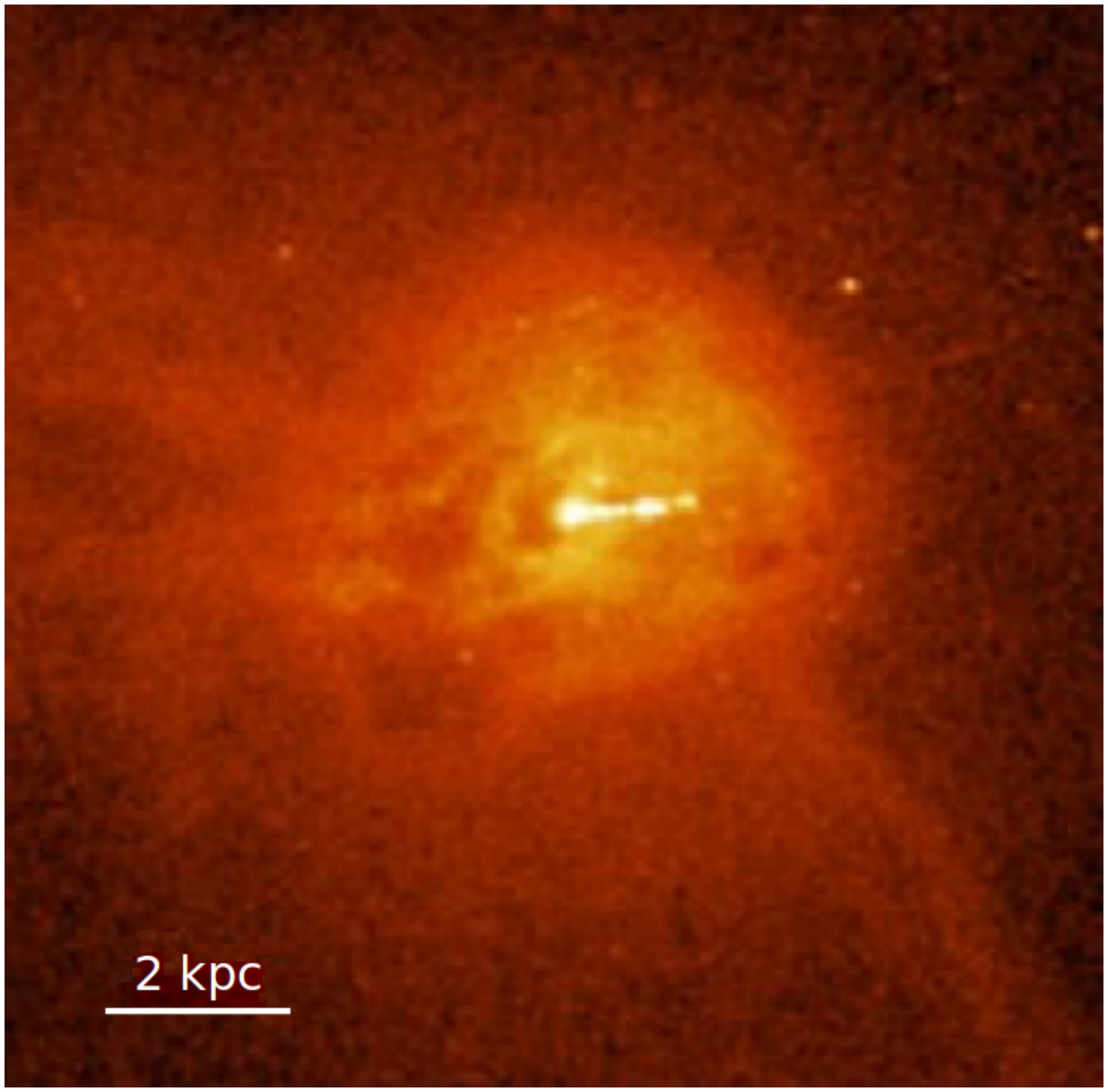}{0.3\textwidth}{M87}
              %\fig{img/MS0735_orig_scale.png}{0.3\textwidth}{MS0735+7421}
              \fig{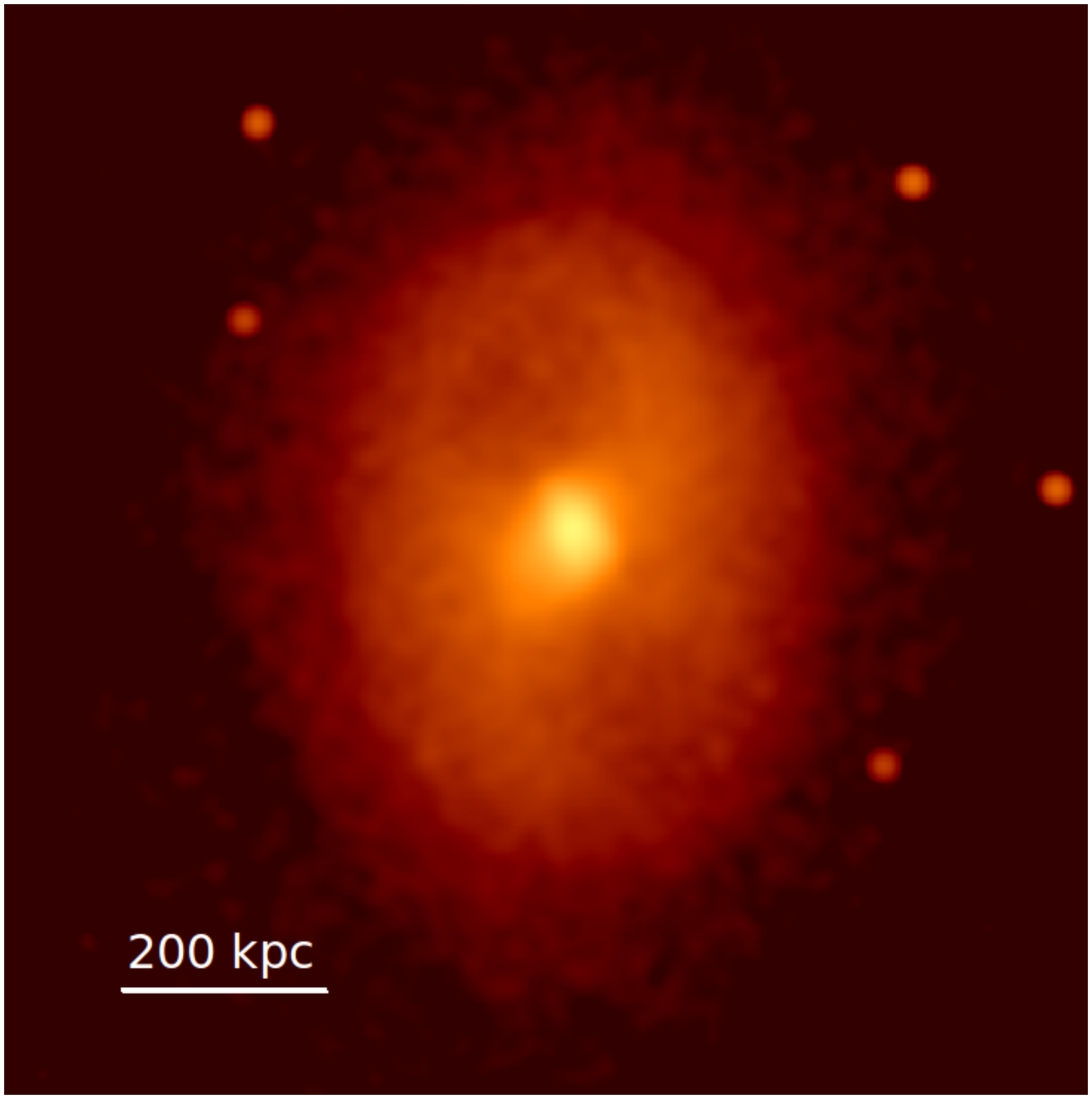}{0.29\textwidth}{MS0735+7421}
    }
	\caption{Chandra X-ray images of Abell 2029 (left), M87 (centre), and MS0735+7421 (right).  Several X-ray cavities are seen in the inner several kpc
	of M87's hot atmosphere \citep{Forman07, Forman16}, and the enormous, 200 kpc diameter cavities are seen in MS0735+7421.  A second pair of cavities in the inner 20 kpc of MS0735+7421, indicating
	very recent AGN activity, are present but not shown here \citep{Vantyghem14}.  No prominent cavities are evident in Abell 2029's hot atmosphere \cite[][]{Paterno-Mahler13}.}
	\label{fig:f3}
\end{figure}

This is not to say that all clusters with multiphase gas fail the $t_c/t_{ff} \lesssim 10$ instability criterion.  But we find systems that obey the criterion do so because their central cooling times are short and not because free-fall times are long. Among the several studies that have examined this criterion \cite[][]{Sharma12,Voit15a,Voit15b,Loubser16}, all did so by calculating the ratio of the average cooling time in radial bins and dividing by an estimate of the free-fall time from a given altitude using a similar approach to ours.  These studies assumed that thermal instability ensues
if and where the minimum value of the $t_c/t_{ff}$ profile falls below 10.   Minima are usually found at a radius of about 10 kpc, consistent with  Figure 1.  Their measured locations depend on several physical and non-physical effects, including the relative slopes of the gas and free-fall time profiles and instrumental resolution effects.  The largest uncertainty concerns the value of the free-fall time, as the acceleration is difficult to measure using standard techniques: stellar velocity dispersions are in short supply, and hydrostatic mass measurements are difficult to measure using standard techniques. The limited number of available velocity dispersions led \citet[][]{Voit15a}, understandably, to calculate the free-fall time adopting a velocity dispersion floor of $\sigma = 250 ~\rm km ~s^{-1}$. While reasonable on average,  this assumption biases $t_c/t_{ff}$ artificially low as stellar velocity dispersions in central cluster galaxies often lie well above $300 ~\rm km ~s^{-1}$. For example, M87 and Abell 2029, with velocity dispersions of $\sigma \sim 340 ~\rm km ~s^{-1}$  \citep{gebhardt11} and $\sigma \sim 400 ~\rm km ~s^{-1}$ \citep{fisher95}, respectively, drive $t_c/t_{ff}$ well above 10 (Figure 1).

That central cooling time is driving the ratio is evident in \citet[][]{Loubser15}, who studied star formation in a sample of 18 central galaxies, four of which are forming stars. The central cooling times of all four star formers lie below 0.5 Gyr, and their ratios of cooling time to free fall time are claimed to lie below ten, i.e., they apparently obey both the $t_c \lae 5\times 10^8$ yr and $t_c/t_{ff} \lesssim 10$ criteria.  The central cooling times and cooling time to free-fall time ratios for the remaining red central galaxies, which are largely devoid of star formation, exceed 1 Gyr and 10, respectively. Thus these objects are consistent with both criteria.  However, the free-fall times for the short and long cooling time systems are consistent with a single value with small dispersion.  The average free-fall time for the star forming systems and dormant systems are  $<t_{ff}> = 0.064 \pm 0.016$ Gyr and $<t_{ff}> = 0.055 \pm 0.014$, respectively.  
%The free-fall times for the red and blue objects are statistically indistinguishable and are consistent with a single value with a small dispersion. 
Therefore, $t_c/t_{ff} $ in this study, as in others, is insensitive to the free-fall time and is governed {\it entirely} by the cooling time, $t_c$.  
We performed an analysis for \citet{Voit15b} and found, similarly to \citet{Loubser15}, that dividing by the free-fall time only increases the scatter in the fundamental relationships between central cooling time, entropy, H$\alpha$ emission \citep{Cavagnolo08}, and star formation \citep{Rafferty08}.  {Therefore, the $t_c/t_{ff} \lae 10$ criterion, as it has been applied, does not indicate the
onset of cooling instabilities.

\subsection{Comparison between Observation and Simulation}

Three dimensional, high resolution simulations of the effects of AGN feedback on cooling atmospheres have offered new
insights into AGN feedback. 
% Nevertheless, observed values of central gas densities and central cooling times and their response to AGN outbursts,
%are at odds with some precipitation models for AGN feedback. 
For example, \citet{Li14}, and
\citet{Li15},  modeled the response of X-ray atmospheres to AGN feedback over a period of several Gyr using an
adaptive mesh refinement code, while \citet{prasad15} used two and three dimensional hydrodynamic simulations. 
Despite significant differences in approach, their model predictions are broadly similar to each other.
In these simulations, hot atmospheres experience large swings in gas density, cooling time,
molecular gas mass, and star formation rate in response to AGN power output over a span of several Gyr.  
The AGN power variations and star formation rates roughly correlate with rising and falling levels of molecular
gas cooling from hot atmospheres. For example,  \citet{Li15}  found swings in the minimum value of the cooling time and minimum value of
$t_{\rm c}/t_{\rm ff}$ that vary by two orders of magnitude and factors of 25, respectively, as
the atmosphere breaths in response to variations in AGN power. \citet{prasad15} found similarly large amplitude swings 
in  minimum $t_{\rm c}/t_{\rm ff}$ and jet power.  The molecular gas mass in these models likewise varies 
by three to four orders of magnitude as it is consumed by star formation.  In the \citet{Li15} model, the molecular gas levels peak when the black
hole, and presumably the radio AGN, are at maximum accretion and power, respectively.  The star formation rate,
jet power, and molecular gas mass all move roughly together, albeit with a lag in time, over the several Gyr simulations.  As the cold gas is
consumed by star formation, the jet power diminishes as its fuel supply subsides.  Declining jet power causes the atmosphere to contract,
the atmospheric gas density rises causing $t_{\rm c}$ to drop as $t_{\rm c}/t_{\rm ff}$ approaches a minimum near unity, and the cooling cycle begins anew.  

While these models capture the quasi-periodic nature of AGN feedback, we are unable to match the observed minimum values of $t_{\rm c}/t_{\rm ff}\simeq 20$ 
MS0735, M87, and Abell 2029 to the predicted minimum cooling time, jet power, molecular gas mass, and star formation rates at any time during the several Gyr spanning the models.    
For their minimum values of $t_{\rm c}/t_{\rm ff}$,
the models generally predict star formation rates, molecular gas, and cooling times at levels exceeding those observed. 
Only Abell 1835, for which we find a minimum $t_{\rm c}/t_{\rm ff}\simeq 10$,  may correspond to an acceptable solution at $\sim 4-5$ Gyr
in the models of \citet{Li15} and \citet{prasad15}, as molecular gas builds up at late times.

Observational trends with molecular gas masses \citep{Edge01} may offer additional insight.   For example, no clear trend between {\it total} molecular gas mass and AGN 
power is found in central cluster galaxies \citep{mcnamara11}.  The relationship
between molecular gas mass and jet power reveals a three decade scatter  in both variables, superposed, perhaps, on a weak trend.  
%The absence of a trend appears at variance with  the \citep{Li15} model prediction, but may be in better agreement with the
%A detailed model comparison is hampered by the inability to determine the time evolutionary state of an individual object, and the observed
%value of $t_{\rm c}/t_{\rm ff}$ does not pin it down. 
%jet power, and molecular gas mass is inconsistent with the data, as  the model $t_{\rm c}/t_{\rm ff}$ values lie well below those observed. 
Our main point is that high molecular gas masses  do not necessarily lead to powerful AGN activity, at least when considering only
central cluster galaxies (this may not be true for lower jet power elliptical galaxies).  This point
is clearly illustrated by the enigmatic MS0735+7421.  At $\sim 10^{46}~\rm erg~s^{-1}$ over the past few hundred Myr, MS0735+7421 is the most
energetic AGN outburst known.  More importantly, its central atmospheric gas density at 10 kpc is similar to others with vastly lower AGN power, and its molecular gas mass
lies well below much weaker AGN.  \citet{prasad15} have argued using their model that the large scatter between molecular gas mass and
jet power arises because the molecular gas is locked-up in kiloparsec-scale disks \citep[e.g.][]{Gaspari12} that are unable
 to fuel the central black hole.  Whether this is generally true is not clear.  
 ALMA observations, which are in short supply, would be required to resolve the molecular gas and to test the molecular disk hypothesis. The main difficulty
 comparing observation to the models is that we are unable 
 to identify a reliable observational marker of the time-evolutionary state of these systems ($t_{\rm c}/t_{\rm ff}$ is unsuited). 
 Once we do, ALMA will in principle test whether objects with the highest molecular gas masses are evolutionarily advanced.

The crux of the problem, in our view,  is
that we do not observe the large amplitude swings in atmospheric gas density (Figure 2) or cooling time (Figure 1) that correspond to
atmospheric ``overheating'' implied by simulation \citep{Gaspari12, Li14, Li15}.   Nor do we find  that $t_{\rm c}/t_{\rm ff}\lae10$ corresponds to the onset of cooling instability.
However,  \citet[][]{Li14, Li15}  pointed out that cooling is enhanced in their simulation by turbulence and when AGN lifted
hot gas to higher altitudes \citep{McNamara14, Voit15b}.  Both processes tend to increase the infall time of the cooling gas driving  the local value $t_c/t_{ff}$ toward  unity. 
This is an important result that we believe is key to understanding thermal instability of hot atmospheres.  }

\section{A Mechanism for Stimulated Feedback}

We propose an alternative mechanism for driving cooling instabilities in hot atmospheres surrounding elliptical galaxies \cite[][]{Werner14} and central cluster galaxies \cite[][]{Edge01,O'Dea08} that incorporates essential physics of the \citet[][]{McCourt12} and \citet{Sharma12} model and the precipitation model of \citet[][]{Voit15a}, but is motivated primarily by observation. 
The mechanism, which we refer to as stimulated feedback, simply posits that molecular clouds condense from cool, low-entropy gas lifted in the wakes of buoyantly-rising X-ray bubbles
to an altitude where the timescale for the clouds to return to their equilibrium position in the central galaxy approaches their radiative cooling time, i.e., $t_c/t_I \lae 1$. Here $t_I$ is the
infall time for thermally unstable clouds whose value depends on factors that may vary within and among systems.
The dynamics of a cloud are determined by, at least, the competing effects of gravity and drag  \citep[e.g.][]{CFN80,nulsen86, Pizzolato05}, and flow in the hot gas.  If the terminal speed of the cloud is smaller than typical infall speeds, it can be lifted and pushed around by hot gas flows and its infall speed will not generally exceed its terminal speed, $\simeq v_K \sqrt{r \delta\rho / (R \rho_e)}$, where $R$ is distance to the cluster center, $\rho_e$ is the ambient gas density, the cloud density is $\rho_e + \delta\rho$ and its depth is $r$.  If the terminal speed of the cloud is greater than typical infall speeds, the cloud will free-fall.  Thus the infall speed of a cooling cloud is generally limited to the lesser of its terminal speed and the free-fall speed.  Angular momentum, magnetic and other stresses might further complicate cloud dynamics, but they do not alter this conclusion.

% {\bf The dynamics of a cloud are determined
%by the competing effects of gravity and drag \citep[e.g.][]{nulsen86, Pizzolato05}, and flow in the hot gas.  If
%the terminal speed of the cloud is smaller than typical infall speeds,
%it can be lifted and pushed around by hot gas flows and its infall
%speed will not generally exceed its terminal speed.  If the terminal
%speed of the cloud is greater than typical infall speeds, the cloud
%will free-fall.  Thus the infall speed of a cooling cloud is generally
%limited to the lesser of its terminal speed and the free-fall speed.
Molecular cloud speeds observed with ALMA \cite[][]{McNamara14,David14,Russell16} indicate that the infall timescale is likely a few
times longer than $t_{ff}$. This mechanism is motivated by ALMA and Chandra X-ray observations of molecular and atomic gas in central galaxies
% \cite[][]{McNamara14,David14,Russell16, Kirkpatrick11}  
indicating hot and cold gas flows behind buoyantly-rising X-ray cavities, and by insights from numerical simulations that closely resemble observed molecular cloud 
morphologies \cite[][]{Li14,Brighenti15}. In the next section we describe the observational indications for this mechanism.
%and  thermal instability models \cite[][]{nulsen86}.

% and their surprisingly slow speeds, while incorporating the
% salient physics of $t_c/t_{ff}\lae 10$ prescription of Sharma et al. (20XX) and McCourt et al. (20XX).  

\subsection{Driving Molecular Gas Flows by AGN in Clusters \& Groups}

Studies of the Perseus cluster have revealed an association between molecular clouds, H$\alpha$ filaments, and buoyantly-rising X-ray bubbles \cite[][]{Salome11}.  Tendrils of molecular gas have apparently been lifted tens of kpc in altitude from NGC 1275 into the Perseus cluster with velocity fields consistent with inflow and/or outflow. 
%of nearly $10^{10}~M_\odot$ of molecular gas. 
ALMA observations of several other central cluster galaxies (including Abell 1835 and PKS 0745-191) have since revealed similar molecular cloud complexes lying beneath buoyantly-rising X-ray cavities and in thin filaments, with velocity fields and locations that further suggest a close relationship between X-ray bubbles and molecular gas \cite[][]{McNamara14,Russell16}.  %Observation indicates that molecular gas flows driven by radio AGN is maybe a common phenomenon with significant implications for galaxy formation.

The Cycle 0 observations of  Abell 1835 \cite[][]{McNamara14} revealed two velocity components.  A fast component of $10^{10}~M_\odot$ of molecular clouds traveling with projected speeds of $200-500 ~\rm km~s^{-1}$ , and a slowly-moving, $4\times 10^{10}~M_\odot$ component of molecular gas. The fast component lies beneath buoyantly-rising X-ray cavities at altitudes
of $5-10$ kpc, while the slow component lies at the center  of the galaxy associated with $\sim 200 ~M_\odot ~\rm yr^{-1}$ of star formation.  The fast clouds are likely an outflow propelled by the rising X-ray bubbles. However, it is unclear how the molecular clouds are accelerated.  Are molecular clouds themselves lifted and accelerated by the bubbles and jets, or is molecular gas condensing from hot, keV gas lifted in the bubbles' wakes?  The answer may be ``both." Acceleration is an issue not just in clusters but in active galaxies in general \cite[][]{Morganti15,Cicone14}, where the molecular outflows are thought to be driven by jets \cite[e.g.,][]{nesvadba08,Wagner11} and winds.  

The difficulty lifting $10^{10} ~M_\odot$ of molecular clouds by jets and bubbles has been discussed in detail by \citet[][]{Russell16}, \citet{McNamara14}, and \citet[][]{David14}, who found in all cases that AGN release enough energy to account for the kinetic energy in molecular gas flows.  However,  whether low density jets have sufficient momentum to accelerate molecular clouds, whose densities exceed jet densities by four or five orders of magnitude, is unclear. 

The total momentum flux (force) available from a kinetic energy dominated jet of power $P_j $ is $(\Gamma_j + 1) P_j /  (\Gamma_j v_j$), where $v_j$ is the jet speed and $\Gamma_j$ is the corresponding Lorentz factor. The buoyant force due to a bubble is $F_B = \rho_e gV$, where $g = v_K^2 /R$ is the acceleration due to gravity. Estimating the volume as $V = H/(4p) = P_j t_j/(4p)$ gives $F_B = (P_j/v_B)[v_K^2 /(4p/\rho_e)]$, where the mean speed of the bubble as it formed is $v_B = R/t_j$, and $H$ is the bubble's enthalpy. The factor in square brackets is of order unity and the mean speed of the bubble is comparable to the sound speed, so the buoyant force exceeds the jet ram pressure unless the flow speed of the jet is transonic relative to the atmosphere or slower. The small cross sections of jets makes them even less effective at lifting.  

Bubble buoyancy is generally more effective at lifting than the ram pressure of the jet that inflated it.  
The lifting ability of radio bubbles in general is limited by Archimedes' principle which prohibits them from lifting more weight than they displace.  The displaced mass in Abell 1835 is uncomfortably close to the $\sim 10^{10}~\rm M_\odot$ of molecular gas flowing behind the bubbles \cite[][]{McNamara14}, while in PKS 0745-191 the bubbles displace roughly ten times the molecular gas mass \cite[][]{Russell16}. However, \citet[][]{David14} found that the molecular gas mass substantially exceeds the  displaced mass in NGC 5044. Therefore,  bubbles may be able to lift the molecular gas in Abell 1835 and PKS 0745-191,  while NGC 5044's bubbles would be unable to do so.  

Additional clues may be found from the radio sources themselves.  Observations of the Abell 2597 and  Abell 1795 central galaxies have shown that their radio jets bend by roughly 90 degrees at the locations of molecular clouds and knots of star formation \cite[][Tremblay {\em et al.} in preparation]{mcnamara96, Salome04}. The sharp bending suggests a collision between the ensemble of molecular clouds and jets halted the jets' forward momenta \citep{mcnamara96}.   Therefore, the jets are unlikely to be accelerating the molecular clouds appreciably, at least over the 
time the jets have been in contact with the molecular clouds.  In fact the molecular clouds in Abell 2597 are moving below the circular speed at their radius (Tremblay {\em et al.} in prep.) and should then be falling in rather than being driven out.  

%A similar conclusion was reached by Morganti et al. (2015) for the Seyfert XXXX, which also has
%revealed molecular gas in the vicinity of it's radio lobes.  

In summary, observation indicates that some molecular gas is lifted directly by radio jets and bubbles.  But the often large molecular gas masses relative to the hot gas mass displaced by
X-ray bubbles suggests some or most is lifted in the hot phase.  Lifting hot, volume-filling gas behind the bubbles is easier, and the coolest gas may be able to condense into molecular clouds on the same timescales bubbles rise to their observed locations \cite[][]{McNamara14,Russell16}. 
%The molecular gas may then have condensed from hot gas lifted in the updraft of the bubbles.
%This scenario receives strong support from Chandra observations of hot outflows and the surprisingly low molecular gas velocities seen with ALMA, indicating that molecular clouds condensed %recently and have not virialized in their host galaxies.

\subsection{Does Molecular Gas Condense from Hot Outflows?}

Chandra has revealed columns of high metallicity gas along and behind X-ray bubbles in clusters.  Gas with high metallicity approaching and sometimes exceeding the
Solar value, enriched by stellar evolution, accumulates around central galaxies.  The metal-rich columns of gas extending tens to hundreds of kpc in elevation are thought to trace hot flows lifted outward by radio bubbles \cite[][]{Simionescu08,Kirkpatrick09,Werner10,Kirkpatrick11}.   This phenomenon is also seen in hydrodynamic simulations where metal enriched gas in the central galaxy is propelled outward in the updraft of rising bubbles \cite[][]{Pope10,Gaspari11}. Estimated flow rates of several tens of solar masses per year would be sufficient to account for the observed molecular gas masses. The altitudes achieved are roughly proportional to the square-root of the jet power \cite[][]{Kirkpatrick15,Morsony10} and are on the order of tens of kpc for typical cluster AGN, but several hundred kpc in the most powerful systems such as MS0735+7421 and Hydra A \cite[][]{Kirkpatrick11,Simionescu08}.  Therefore the lifting of hot gas by radio bubbles would be at least a plausible source of fuel for molecular cloud condensations at high altitudes.
%To drive the system to thermal instability the cool gas in the centre must be lifted to an altitude where it's cooling time is equal to a few times the cloud infall time. 

%\subsection{Molecular Clouds Condense from Hot, Metal-enriched Outflows}

Among the puzzling results from early ALMA observations of clusters are the surprisingly low radial velocities and velocity widths of the molecular gas. In response to the slow inner cloud velocities in Abell 1835, we suggested the mundane and unlikely possibility that the slow (radially) moving nuclear clouds are rotationally supported in a disk viewed in the plane 
of the sky \citep{McNamara14}. While this interpretation may apply to Abell 1835 itself, cloud velocities lying well below the stellar velocity dispersion or circular speeds have now been seen in several systems \cite[][Tremblay {\em et al.} in prep.]{David14,Russell15,Russell16} indicating a phenomenon unrelated to rotational support or orientation effects.  Traveling well below their expected gravitational speeds, either the clouds have had no time to relax in their gravitational potential wells and thus are surprisingly young, or they are pinned by magnetic fields or dynamic pressure to the hot gas from which they cooled, or some combination \cite[][]{Russell16}. Regardless of the cause, if molecular cloud velocities are representative of the speeds of condensing clouds, their velocities are several times lower than the
their hosts' stellar velocity dispersions and below their expected free-fall speeds.  %Slower infall speeds of thermally unstable clouds would drive the ratio of the cooling time to the {\it infall } time toward unity, and could enhance cooling behind the bubbles.

\subsection{Stimulated Cooling at High Altitudes}

Following the arguments of \citet[][]{nulsen86}, low entropy, $\lesssim \rm 1~keV$ gas should condense into molecular clouds when lifted to an altitude where the ratio of its cooling time to {\it infall} time, $t_c/t_{I}$, approaches unity. Here, the infall time can be longer than the free-fall time as indicated by ALMA observations.  
We postulate that systems with short central cooling times $t_{c}\lesssim 5\times 10^8~\rm yr$, yet lacking X-ray cavities powerful enough to lift  low entropy gas to altitudes where $t_c/t_{I}\sim 1$, remain thermally stable and thus do not shine with nebular emission.   

For example, Abell 2029 is apparently thermally stable i.e.,  $t_c/t_{I}> 1$ throughout its hot atmosphere.  Figure 1 shows that the mean atmospheric cooling time within
10 kpc is $t_c = 3-5 \times 10^8~\rm yr$.  Over this volume, $t_c/t_{ff} = 20-30$ and its atmosphere remains thermally stable, despite its short central cooling time. The atmosphere will become unstable where $t_c/t_{ff}\lae 1$. Its AGN must then lift hot gas from the inner 10 kpc to altitudes between $\sim 280-400$ kpc
where the free-fall time is approximately equal to the cooling time of the low entropy gas lifted from within 10 kpc.  
Relaxing the thermal instability criterion to $t_c/t_{ff} \lae 10$ implies lifting altitudes between $15-30$ kpc.  Upon lifting the cooler, denser central
gas into the lower pressures atmosphere at higher altitudes, the lifted gas will expand and cool.  Being denser than its surroundings, it should detach from the outflowing gas behind the bubble
and fall back to the galaxy as it condenses into molecular clouds.   If the cooling gas remains tethered 
to the surrounding atmosphere by magnetic fields or dynamical pressure as ALMA observations suggest, its infall time, being governed by the terminal speed, 
would exceed the free-fall time.   This would reduce the lifting altitude required to initiate cooling, such that $t_c/t_{I}$ approaches unity.  In this picture, Abell 2029 fails
to shine in H$\alpha$ emission because its radio emission,  despite being fairly powerful, lies within 25 kpc of the nucleus \cite[][]{Paterno-Mahler13}.  Abell 2029 has apparently not developed 
cavities capable of lifting low entropy gas to the altitudes required to destabilize it.

A similar analysis for MS0735+7421
gives similar figures.  However, unlike Abell 2029, its powerful X-ray bubbles are lifting hot gas to altitudes upward of 300 kpc \citet{Kirkpatrick11}, well beyond the elevations
required to stimulate cooling.  As expected, its H$\alpha$ emission extends $20-40$ kpc in the central galaxy with a luminosity among the highest known in a cluster \cite[][]{Donahue92}.
The third example, M87, lying in a cooler atmosphere, will destabilize at lower altitudes, and thus requires a lower power AGN to stimulate cooling.  Its average cooling time within a kpc or so lies below  $\sim 10^8~\rm yr$.  Therefore following on the previous examples, its bubbles must lift this gas to altitudes of only $5-20$ kpc or so to initiate thermal instability. Shock fronts, bubbles \citep{Forman07} and metal-enriched  gas columns \cite[][]{Simionescu08} are observed in M87 to elevations exceeding 10 kpc,
and nebular emission is observed within a similar volume \citep{Sparks04}, which again is consistent with our model.

Perhaps the best example is the Perseus cluster which contains $\sim 10^{10} ~M_\odot $ of molecular gas centered on NGC 1275 and in filaments extending to altitudes of $30-50$ kpc \cite[][]{Fabian03,Salome11}.  The cooling time of its ambient gas exceeds the free-fall time by more than an order of magnitude and should be thermally stable as $t_c/t_{ff} >10$ at these elevations.  However the association between molecular gas, H$\alpha$ filaments, and its system of X-ray bubbles is consistent with low entropy gas lifted to altitudes where it can cool and return to fuel star formation in NGC 1275.

Finally, we have examined archival Chandra images of the five ``spoilers" in \citet[][]{Cavagnolo08}, those systems whose central atmospheric cooling times fall below the cooling time threshold yet lack detectable H$\alpha$ emission. Like Abell 2029, and shown here in Figure 4, only one, RBS0533 has revealed a possible cavity located approximately 10 kpc to the south-east of
its centroid, but otherwise no prominent cavity systems are seen (Abell 2029 is among the five).  The archival images shown in Figure 4 have not been exposed deeply enough to exclude
possible faint cavities at large radii.  Nevertheless, until deeper exposures are obtained, the data in hand are consistent with our phenomenological model.  Stimulated feedback is thus a viable and testable alternative to the
conceptually important thermal instability \citep{McCourt12} and precipitation \citep{Voit15a} models, but which are inconsistent with observation.

%While some molecular gas may be lifted by the jets and or bubbles, the data indicate most of the lifting is more easily and likely accomplished in the hot phase. The situation in groups and clusters differs qualitatively from observations of quasars and Seyferts (Morganti et al. 2015, Cicone et al. 2014), where the molecular outflows are less massive, on a smaller scale, and thus may be driven by jets (eg. Wagner \& Bicknell 2011) and winds.  

%\begin{figure}[htbp] %  figure placement: here, top, bottom, or page
%   \centering
%   \figurenum{2}
%	\gridline{\fig{img/A1835_molecbehindbubbles.jpg}{0.3\textwidth}{}
%   }
%   \caption{ALMA map of molecular clouds trailing rising X-ray cavities (indicated by circles) from McNamara et al. (2014)}
%  	\label{fig:f2}
% 	\end{figure}

\begin{figure}[htbp] %  figure placement: here, top, bottom, or page
   \figurenum{4}
   \centering
	\gridline{\fig{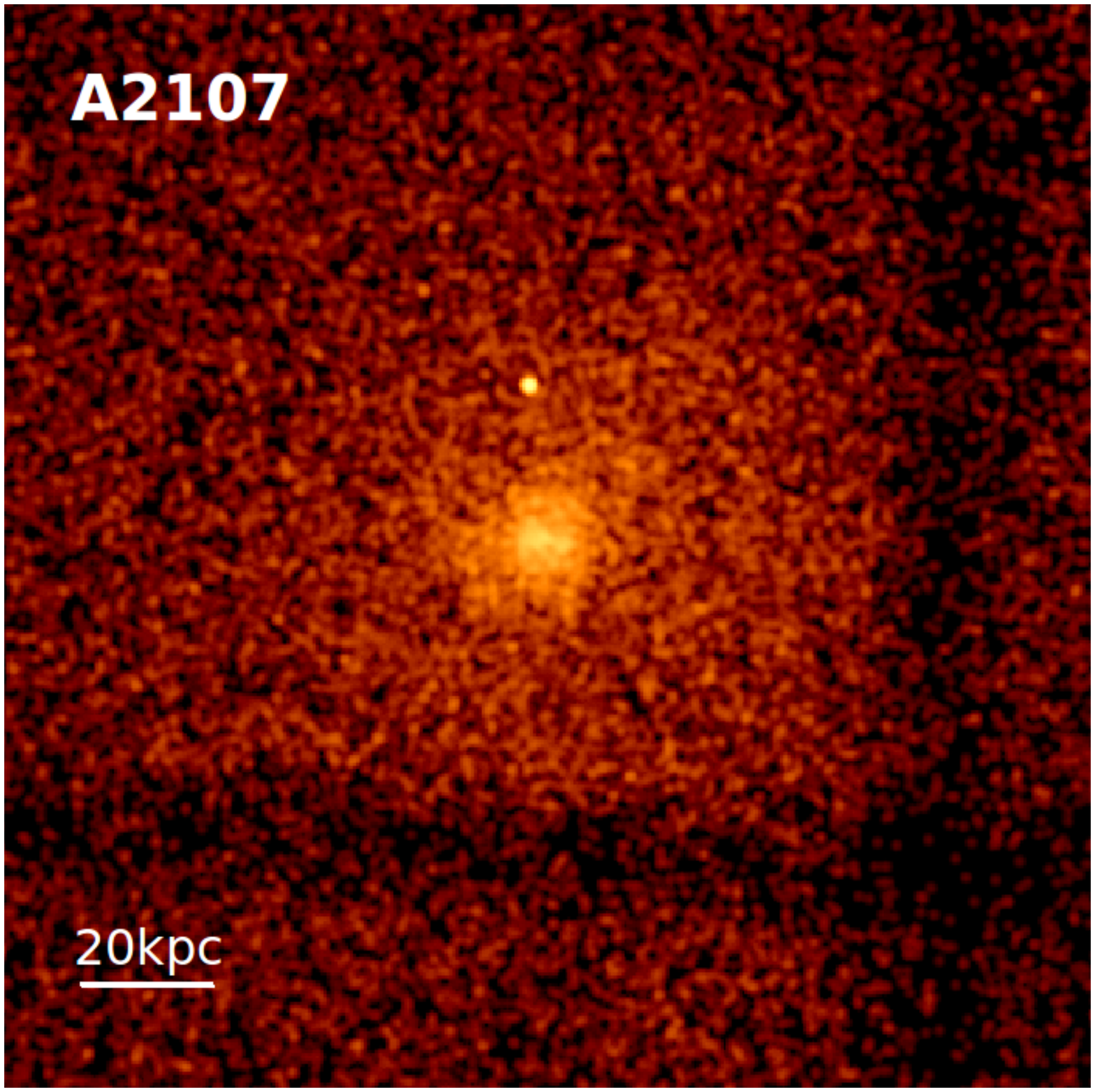}{0.25\textwidth}{}
             \fig{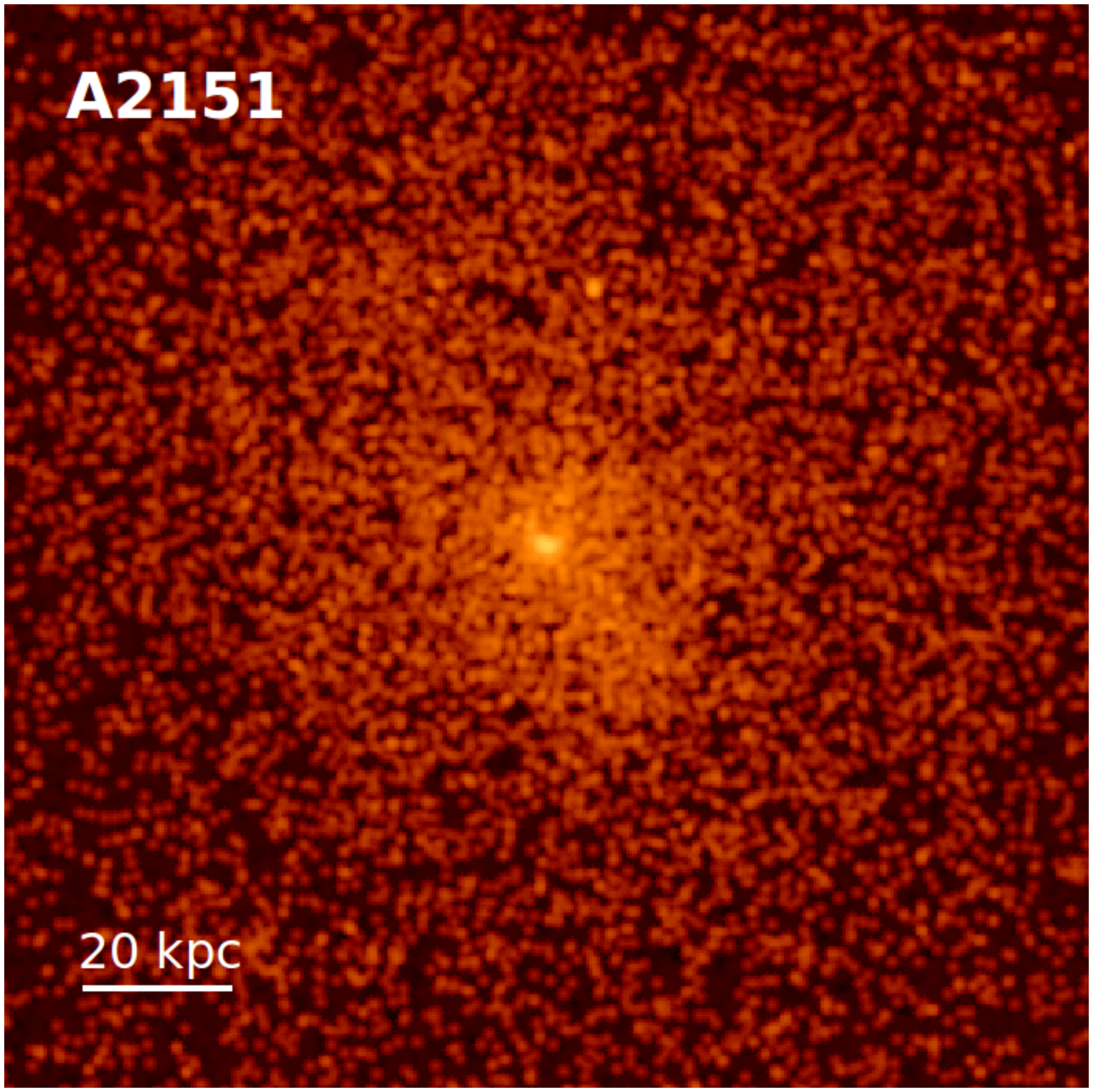}{0.25\textwidth}{}
             \fig{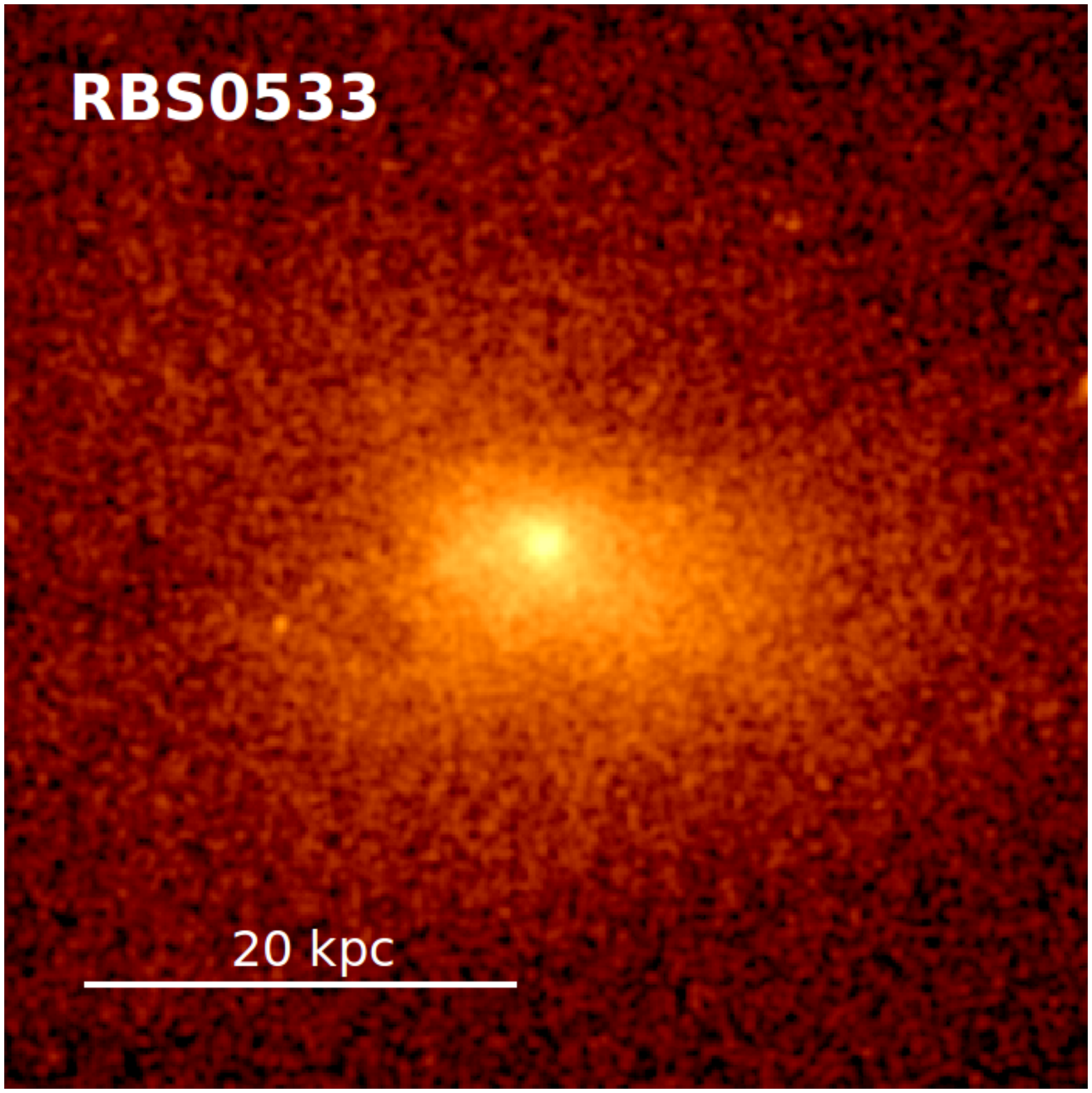}{0.25\textwidth}{}
             \fig{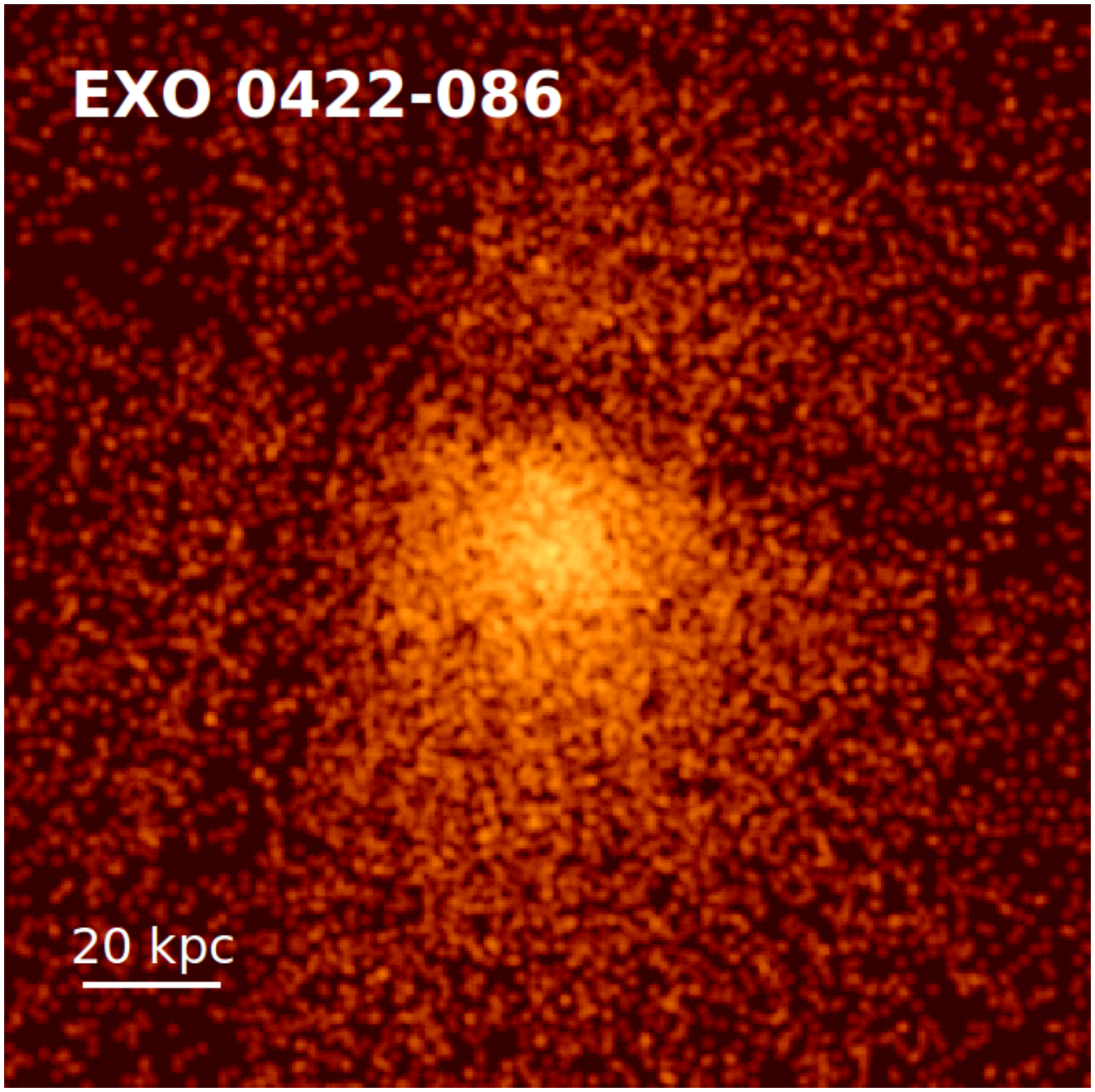}{0.25\textwidth}{}
}
 \caption{Chandra X-ray postage stamp images of the central 200 by 200 arcseconds of the ``spoiler" clusters from  Cavagnolo et al. (2008): Abell 2107, Abell 2151, RBS0533, EX00422-086.  The fifth spoiler is Abell 2029, shown in Figure 2.  None reveal prominent radio bubbles.}
   \label{fig:f4}
   \end{figure}

\section{Summary and Final Remarks}

Motivated primarily by new ALMA observations, we have proposed a new phenomenological model for the onset of thermal instabilities leading to nebular emission and
star formation in giant galaxies.  Molecular condensations form when low entropy gas lying within a central galaxy is lifted to higher altitudes behind buoyantly-rising X-ray bubbles inflated by radio AGN.  Lifting the low entropy gas effectively increases  its {\em infall} time, promoting condensation into molecular clouds in the bubbles' wakes.  The surprisingly slow molecular cloud velocities found by ALMA indicate that the infall timescale, $t_I$, can be significantly longer than the free-fall timescale.
In this new picture, the rising X-ray bubbles responsible for heating hot atmospheres simultaneously lift molecular gas and promote cooling in their wakes, fuelling a steady feedback loop in a mechanism we refer to as stimulated feedback.  Molecular clouds eventually return to the central galaxy in a circulating flow that fuels star formation and the AGN itself.  Once stimulated feedback commences it naturally sustains itself, consistent with the prevalence of feedback in clusters and galaxies to large look-back times \cite[][]{Best07,Ma13,Hlavacek-Larrondo15}.    

The amount of hot gas available to be lifted within the inner $10-20$ kpc of most clusters lies between several $10^9\rm ~M_\odot$ to several $10^{10}\rm ~M_\odot$, which would amply supply the observed levels of molecular gas in most galaxies \cite[][]{Edge01}.  In extreme instances, such as Abell 1835, where the molecular gas mass in the central galaxy 
%is $\sim 5\times 10^{10}\rm ~M_\odot$ \cite[][]{McNamara14} 
exceeds the hot gas mass within a similar volume, the fuel supply must have accumulated from multiple AGN outbursts or have been augmented by other cooling channels.  
%We have suggested that thermal instabilities leading to the condensation of molecular clouds, nebular emission, and in some instances, star formation \cite[][]{Donahue15} are triggered in the updraft of rising radio bubbles, that at the same time, are stabilizing hot atmospheres against large-scale cooling. 
%Our model resolves the observational difficulties with the \citet{McCourt12}, \citet{Sharma12}  and closely-related precipitation models \citet{Voit15a}.  
Our phenomenological model can be ruled out if it can be shown to be inconsistent with the cooling time and entropy thresholds.  It must explain systems lying below the thresholds that lack H$\alpha$ emission,
which implies that their radio AGN are too weak to lift gas to an altitude where it becomes thermally unstable.  The model has the interesting property that once it gets started it is potentially
self-sustaining, which is an essential aspect of any feedback model.  Because in stimulated feedback the jet must be able to lift the gas that eventually cools into molecular
gas and stars to high altitudes, it may be more stable and less prone to overcooling that leads to unrealistically
high molecular gas masses and star formation rates seen in precipitation simulations.

Understanding how stimulated feedback leads to the precise values of the cooling time and entropy thresholds for the onset of  H$\alpha$ emission and star formation, and what the value of $t_c/t_I$ must be to stimulate thermal instability, are interesting challenges.  That the cooling time threshold's value of $t_c \sim 5\times 10^8~\rm yr$ is close to the maximum cycle duration for AGN feedback \cite[][]{Birzan13,Vantyghem14} is noteworthy.  Furthermore, the lifting altitude at which the free-fall time
is roughly equal to the value of the cooling time threshold is several hundred kpc in the most massive clusters.  This altitude is close to the highest lifting altitudes achieved by powerful AGN in clusters \citep{Kirkpatrick15}. Therefore, the cooling time threshold may be set by the jet power itself.  It must also be understood why central galaxies bright with H$\alpha$ emission that obey the cooling time threshold do not all have appreciable ongoing star formation.  Those include the three objects highlighted here. We suggested that cooling instabilities in the star forming galaxies have advanced to the point that they have accreted the critical surface density of molecular clouds required for stars to form. 
Finally, the question of what gets the mechanism started need not be a problem.  An influx of gas, whether from the X-ray atmosphere or
stripped from a passing galaxy or merger, could initiate it.  These questions can be tested 
observationally and explored theoretically, and may eventually overcome the problems with precipitation models.
% \citet{McCourt12}, \citet{Sharma12}, and closely-related precipitation model \citep{Voit15a}.

\bigskip

BRM acknowledges generous financial support from the Natural Sciences and Engineering Research Council of Canada and the Canadian Space Agency. 
HRR and ACF acknowledge support from ERC Advanced Grant 340442. This work was supported in part by Chandra Award Number G05-16134X. We acknowledge
helpful discussions with Mark Voit, Megan Donahue, Prateek Sharma, and the anonymous referee.

\appendix

\section{M87}

Owing to M87's proximity, we are able to follow its declining cooling time profile in Figure \ref{fig:f1} from a value of $\sim 1 \rm ~Gyr$ at an altitude of 20 kpc to $3\times 10^7 ~\rm yr$ within 200 pc of the nucleus \cite[][]{Russell15}. The atmospheric cooling time remains near to or below $10^8\rm ~yr$ within 1 kpc, well below the cooling time threshold. Several X-ray cavities and a series of weak shock fronts lie within 10 kpc of M87's nucleus associated with its radio source, indicating a total AGN power of $8\times 10^{42} ~\rm erg~s^{-1}$ \cite[][]{Forman07,russell13}. M87 harbours bright nebular emission within 10 kpc where the the cooling time lies below $t_{c}=6\times 10^8~\rm yr$, consistent with the cooling time threshold. 

\section{MS0735+7421}

The MS0735+7421 cluster's AGN, the most energetic known,  has inflated enormous cavities, 200 kpc in diameter, \cite[][]{mcnamara05} with total energy expended by its cavities and surrounding shock fronts approaching $10^{62} \rm ~ erg$ \cite[][]{Vantyghem14}. 
%MS0735+7421 contains two cavities, each approximately 200 kpc in diameter indicating a jet power of $10^{46}~\rm erg~s^{-1}$. I
%t is the most powerful AGN known \cite[][]{Vantyghem14}.
While the central galaxy shows no indication of star formation ($<0.5 ~M_\odot ~\rm yr^{-1}$), it contains bright nebular emission indicating cooling, multiphase gas \cite[][]{Donahue92}.
Its H$\alpha$ nebula, with a luminosity of $\sim 10^{42}~ \rm erg~s^{-1}$ extending to 30 kpc in altitude, is among the most luminous known in a galaxy cluster \cite[][]{Donahue92}.  
As expected,  MS0735+7421's atmospheric cooling time within 10 kpc drops to $t_c = 5\times 10^8 ~\rm yr$, which lies close to the cooling time threshold.  

\section{Abell 2029}

We are able to measure Abell 2029's cooling time profile to an inner radius of 5 kpc, where its radiative cooling time approaches $2\times 10^8~\rm yr$, well below the cooling time threshold. In this respect, its X-ray atmosphere is similar to other clusters whose central galaxies are burgeoning with star formation fueled by reservoirs of upwards of $10^9~M_\odot$ of molecular gas, such as  Abell 1795 and Abell 1835, yet it shows no sign of star formation or nebular emission \cite[][]{Johnstone87,Johnstone88,McNamara89}. Abell 2029's central galaxy hosts a strong radio source, with
a 1.4 GHz luminosity $L_{1.4}\sim 10^{42}~\rm erg~s^{-1}$ \cite[][]{Cavagnolo08}. An early claimed detection of weak X-ray cavities \cite[][]{Rafferty06} is not confirmed in deeper X-ray data, although a cold front is visible in Figure 3. Apart from the cold front and some larger scale structure, its atmosphere is relatively smooth \cite[][]{Paterno-Mahler13}.

%Molecular clouds not uniquely associated with star formation but are dusty (Russell et al. 2014, David et al. 2014).Dragging out dust with the X-ray gas may accelerate cooling along the rising trajectory.

%General phenomenon:  Morganti, HZRGs etc.
%Bla... Bla... Bla...

%%%%%%%%%%%%%%%%%%%%%%%%%%%%%%%%%%%%%%%%%%%%%%%%%%



\bibliographystyle{apalike}
\bibliography{refs} % if your bibtex file is called refs.bib

%% Alternatively you could enter them by hand, like this:
%% This method is tedious and prone to error if you have lots of references
%\begin{thebibliography}{99}
%\end{thebibliography}

%%%%%%%%%%%%%%%%%%%%%%%%%%%%%%%%%%%%%%%%%%%%%%%%%%

%%%%%%%%%%%%%%%%% APPENDICES %%%%%%%%%%%%%%%%%%%%%

\end{document}